\documentclass[showpacs,preprint,preprintnumbers,amsmath,amssymb,a4paper,aps]{revtex4}

\usepackage{amssymb}
\usepackage{epsfig}

\begin{document}

\title{A fast simulator for polycrystalline processes with application
to phase change alloys}

\author{Peter Ashwin$^1$, Patnaik BSV$^2$ and C. David Wright$^1$}
\affiliation{$^1$ School of Engineering, Computer Science 
and Mathematics, University of Exeter, Exeter EX4 4QF, UK\\
$^2$ Department of Applied Mechanics,
Indian Institute of Technology Madras, Chennai, 600 036, India}

\begin{abstract}
We present a stochastic simulator for polycrystalline phase-change
materials capable of spatio-temporal modelling of complex anneals. This is based
on consideration of bulk and surface energies to generate rates of
growth and decay of crystallites built up of `monomers' that themselves may be
quite complex molecules. We perform a number of simulations
of this model using a Gillespie algorithm. The simulations are performed at molecular scale and using an approximation of local free energy changes that
depend only on immediate neighbours. The sites are on a lattice chosen to have a lengthscale of
the individual monomers, where each site gives information about a two-state 
local phase $r$ ($r=0$ 
corresponds to amorphous and $1$ corresponds to crystalline) and a
continuous crystal orientation $\phi$ at each site.

As an example we use this to model crystallisation in chalcogenide GST ($GeSbTe$) alloys used for example in phase-change memory devices, where reversible 
changes between amorphous and crystalline regimes are used to store 
and process information. We use our model to simulate anneals of GST
including ones with non-trivial spatial and temporal variation of 
temperature; this gives good agreement to experimental incubation times at low temperatures while modelling non-trivial crystal size distributions and melting dynamics at higher temperatures.
\end{abstract}

\date{\today}

\pacs{
07.05.Tp (Computer modeling and Simulation),
64.60.De (Statistical mechanics of model systems; Ising model, Potts model, field-theory models, Monte Carlo techniques, etc).
}

\maketitle

\section{Introduction}

This paper considers a model for phase change materials, i.e. alloys that can
undergo reversible phase changes in response to anneals.
Possible modelling techniques for polycrystalline processes in reversible phase-change alloys such as GST ($GeSbTe$) range from molecular dynamic simulations at one end of the spectrum to empirical models at the other. The former are thermodynamically realistic but highly computer-intensive; the latter are fast, but hard to relate to material properties. Hence models used in practise tend to lie somewhere between the two, for example, Monte Carlo simulations \cite{MC_sims}, JMAK and master equation based models \cite{Ka1,KGT,SW_2004,WAB_APL} or probabilistic cellular automata \cite{pcaref,YONA,Hyot&al}.

We describe the material as a 2D lattice of discrete `sites' where each site is either crystalline or amorphous and there is an underlying orientation that varies continuously; these sites are on the lengthscale of monomers, though they do not correspond directly to individual monomers. As such our method for generating complex anneals, combines elements of probabilistic cellular automata (PCA) models \cite{pcaref,YONA,Hyot&al}, polycrystalline phase field models \cite{KarRap,CX,VP_2006} and uses thermodynamics from master equation models \cite{Ka1,KGT}. As we use a Gillespie algorithm for time-stepping we refer to it as a `Phase Field Gillespie' model. In particular we retain a high level of simplicity because of discrete time and lattice space model, while retaining thermodynamic realism and hence keep fitting parameters to a minimum.

Recall that crystallisation can be thought of as a two-stage process; nucleation (where a small crystallite needs to overcome an energy barrier dominated by interfacial energy) and growth (where the crystallite grows according to the availability of neighboring monomers and dominated by bulk energy). We assume each site has a set of locally-determined rate constants for transitions into a new state. These rates depend only on the current state of the site and that of its immediate neighbours. For the rates of growth and dissociation for GST we use thermodynamics from \cite{SW_2004,WAB_APL}.

The Gillespie algorithm \cite{GIL_1977} can be used to simulate the evolution under the assumption that the events are independent, instantaneous and never simultaneous. Each step of the algorithm has two parts; firstly it determines a random time to next event and secondly it determines which event occurs. This enables one to perform fast and physically plausible simulations of a number of crystallisation-related phenomena, including incomplete crystallisation, melting and complex spatio-temporal anneals. As there are many possible events, one must efficiently use data-structures to ensure that the simulations run at a high speed and hence perform simulation of complex anneals in 2D on a standard desktop computer.

After the algorithm for the Phase Field Gillespie simulator is described in Section ~\ref{secPFG}, Section~\ref{secResults} presents the results of some bulk anneals of GST using this simulator, showing that the simulator can model nucleation effects, non-trivial anneals and melting.  We include examples where the temperature depends on space and/or time; one can see a variety of effects and a good quantitative agreement with experimental temperature-dependent incubation times for GST. Finally Section~\ref{secDiscuss} discusses some possible extensions and limitations of the method.

\section{A Phase Field Gillespie (PFG) crystallisation simulator}
\label{secPFG}

We consider a homogeneous (though not necessarily isotropic) material 
in 2D. The state of the material is described on 
a discrete regular lattice of grid points $G$ on the length scale of the
individual monomers. Each site is assumed to be either 
`crystalline' or `amorphous'. More precisely, at each grid point, $(i,j)\in  G$,
the state is described by two quantities:
\begin{itemize}
\item $r_{ij}$ -- a discrete `phase' variable that is either $0$ (amorphous)
or $1$ (crystalline).
\item $\phi_{ij}$ -- a continuous `orientation' variable that varies over some
range $[0,\pi)$ and gives a notional representation for the local orientation of the material.
\end{itemize}
In particular we can determine that two adjacent crystalline sites are within the same crystal if and only if $r_{ij}=r_{kl}=1$ and $\phi_{ij}=\phi_{kl}$.

The model we describe is a stochastic model that estimates
rates of possible local changes to the state of the system
(i.e. changes that affect only one site) and uses a Gillespie algorithm 
\cite{GIL_1977} to evolve the system in time.
A Gillespie algorithm is optimal in that it will generate 
timesteps at a rate corresponding to  the fastest rate that 
requires updating, though it is typically more complex to implement 
than Monte Carlo simulations \cite{MC_sims}. Although there are adaptations of the algorithm
to other contexts \cite{Gil_2001} we use the original version of Gillespie.

We consider the following possible instantaneous events at a site $(i,j)\in G$:
\begin{itemize}
\item {\bf Nucleation} -- The site $(i,j)$ and an adjacent site, originally 
both amorphous, become a crystallite at a rate $C^{\rm nu}_{ij}$.
\item {\bf Growth} -- The site $(i,j)$, originally amorphous, becomes attached to
an adjacent crystal of orientation $\psi$ at a rate $C^{\rm gr}_{ij\psi}$.
\item {\bf Dissociation} -- The site $(i,j)$, originally crystalline, detaches or dissociates from the crystal of which it is a part to become amorphous at a rate $C^{\rm di}_{ij}$, and assumes a random orientation.
\end{itemize}

\subsection{The rate coefficients}

We approximate the rate coefficients for nucleation, growth 
and dissociation ($C^{\rm nu}$, $C^{\rm gr}$ and $C^{\rm di}$) at each grid
point by considering the change in bulk and surface energies
of crystallites adjacent to that site. We define the set of neighbours of $(i,j)\in G$
$$
N_{ij} = \lbrace (k,l)\in G : (k,l) \mbox{ is a neighbour of } (i,j) \rbrace,
\mbox{ and } n_{ij} = |N_{ij}|,
$$
the set of amorphous neighbours of $(i,j)$
$$
N^{am}_{ij} = \lbrace (k,l) \in N_{ij} : r_{kl} = 0 \rbrace,
\mbox{ and } n^{am}_{ij} = |N^{am}_{ij}|,
$$
and finally the set of crystalline neighbours of $(i,j)$ with a given orientation $\psi$ 
$$
N^{or}_{ij\psi} = \lbrace (k,l) \in N_{ij} : \phi_{kl}=\psi ~{\rm and}~ r_{kl} =1 \rbrace, 
\mbox{ and } n^{or}_{ij\psi} = |N^{or}_{ij\psi}|.
$$
Note that $n^{am}_{ij},n^{or}_{ij\psi} \in \{0,\cdots,n_{ij}\}$.

The rates are considered in a similar way to the derivation of master equation rates as in \cite{SW_2004} and outlined below. We assume that `interactions' between neighbours occur at a temperature-dependent rate
$$
R(T)= k_0e^{\left(-\frac {E_a}{k_B T}\right)}
$$
where $E_a$ is the activation energy and $k_B$ is the Boltzmann constant. The 
prefactor $k_0$ is used as a fitting parameter to normalise the results; see
\cite{SW_2004}.

If adjacent sites have an `interaction' we define
$$
\xi(T,A) =\left\{
\begin{array}{l}
\mbox{rate at which a site transforms form amorphous to crystalline,}\\
\mbox{resulting in a change A in surface area of the crystallite}.
\end{array}
\right\}
$$
We assume local thermal equilibrium, meaning that the rate of 
the reverse transformation at an interaction is $\xi^{-1}(T,A)$. This rate varies 
with temperature as the bulk and surface energy vary.

We compute the change in surface area of the crystallites on adding
site $(i,j)$ to a neighbouring crystal of orientation $\psi$ by 
a linear approximation
$$
A= S_m \left[\frac{n_{ij}-2n_{ij\psi}}{n_{ij}}\right]
$$
where $S_m$ is the surface area of a single site. This means that changing
an isolated site in the middle of a crystal of orientation $\psi$ will result
in a change $A=-S_m$ (as $n_{ij\psi}=n_{ij}$), while creating a new crystal
in the middle of a field of amorphous material will result in a change 
$A=S_m$ (as $n_{ij\psi}=0$).

Putting this together (and noting that only by interaction with amorphous neighbours can a site nucleate) we get rate for nucleation that is
\begin{equation}
C^{\rm nu}_{ij} = \left\{\begin{array}{ll}
k_0~e^{\left(-\frac {E_a}{k_B T}\right)}~\frac{n^{am}_{ij}}{n_{ij}} \xi(T,S_{m})&\mbox{if}~ r_{ij}=0 \\
0&\mbox{if}~r_{ij}=1.
\end{array}\right.
\label{eq:rn}
\end{equation}
The growth rate for an amorphous site to join a 
crystalline neighbour with orientation $\psi$ is:
\begin{equation}
C^{\rm gr}_{ij\psi} =  \left\{\begin{array}{ll}
k_0~e^{\left(-\frac {E_a}{k_B T}\right)}~ \xi(T, S_{m}\frac{n_{ij}-2n_{ij\psi}^{or}}{n_{ij}})~
& \mbox{if}~r_{ij}=0 \\
0 & \mbox{if}~r_{ij}=1.
\end{array}\right.
\label{eq:rg}
\end{equation}
Finally, the dissociation rate for a crystalline site to become amorphous is:
\begin{equation}
C^{\rm di}_{ij}=   \left\{\begin{array}{ll}
0& \mbox{if}~r_{ij}=0\\
k_0~e^{\left(-\frac {E_a}{k_B T}\right)}~{ \xi(T, S_{m}\frac{n_{ij}-2n_{ij\psi}^{or}}{n_{ij}})^{-1}}~
& \mbox{if}~r_{ij}=1.
\end{array}\right.
\label{eq:rd}
\end{equation}

\subsection{The PFG Algorithm}

We now detail the operation of the PFG algorithm using the rates above. Initially, the whole domain is assumed to be an 
`as deposited' amorphous state with a random distribution 
of $\phi_{ij}$ values and $r_{ij}=0$, though one can restart the 
algorithm from any given state.

For a square lattice we use eight neighbors, $n_{ij}=8$, weighted according  to their distance from the site. The new state of the site is then given by $r'_{ij}$ and $\phi'_{ij}$, using the stochastic simulation algorithm of Gillespie \cite{GIL_1977} as follows. This simulates up to a
time $T_{\max}$.

\begin{enumerate}
\item 
Start at time $T=0$ with given $r_{ij}$ and $\phi_{ij}$.
\item
Generate rate coefficients for all grid points $C_{ij}^{\rm nu}$ $C_{ij \psi}^{\rm gr}$ $C_{ij}^{\rm di}$ for  nucleation, growth and dissociation 
respectively. We refer to these using a single index $\nu=(i,j,{\rm a})$ where 
${\rm a}\in\{{\rm nu,(\rm gr,\psi), \rm di}\}$.
\item
Compute the sum
$$
a_{0} = \sum C^{\rm nu}_{ij} 
+ \sum_{ij} \left[\sum_{\psi\in\Psi_{ij}}C^{\rm gr}_{ij \psi} \right]
+ \sum C^{\rm di}_{ij} 
$$
where $\Psi_{ij}=\{\phi_{kl}~:~(k,l)\in N_{ij}\}$ is the set of orientations
of neighbours to $(i,j)$.
\item
Generate two independent random numbers $\eta_{1}, \eta_{2}$ uniformly
distributed on $(0,1)$ and compute 
\begin{equation}
d\tau = \frac{1}{a_{0}}\log_{e}\left(\frac{1}{\eta_{1}}\right).
\end{equation}
Increment time to $T = T+ d\tau$. If $T\geq T_{\max}$ then
{\bf stop}.
\item
Identify the event $\nu=(i,j,{\rm a})$ corresponding to
 $(i,j)$, ${\rm a} \in {\rm (nu,gr,di)}$
and $(k,l)$ with $\phi_{kl}=\psi$ such that
\begin{equation}
\sum_{\nu =1}^{\mu -1} {a_{\nu}} < \eta_{2}a_{0} \leq
\sum_{\nu =1}^{\mu} {a_{\nu}} 
\end{equation}
\item
Update the value of $\phi_{ij}$  and $r_{ij}$. More precisely,
perform the following updates according to corresponding reactions
(nucleation, growth or dissociation) that occur:
\begin{enumerate}
\item {\bf Nucleation} at $(i,j)$; pick a $(k,l) \in N_{ij}^{am}$ at random and set
$$
\phi'_{ij}=\phi'_{kl}=\phi_{ij}, ~~r'_{ij}=r'_{kl}=1.
$$
\item {\bf Growth from neighboring crystal} with $\psi=\phi_{kl}, (k,l) \in N_{ij}$ into the amorphous site $(i,j)$; set
$$
\phi'_{ij}=\phi_{kl}, ~~r'_{ij}=1.
$$
\item {\bf Dissociation} at $(i,j)$; set
$$
r'_{ij}=0~~  \phi'_{ij} = W,
$$
where $W$ is an independent random number uniformly distributed in the range of possible orientations $[0,\pi)$.
\end{enumerate}
\item
For the next iteration, copy $\phi_{ij}=\phi'_{ij}$ and $r_{ij}=r'_{ij}$ and update the values of  $C_{ij}^{\rm nu}$ $C_{ij \psi}^{\rm gr}$ $C_{ij}^{\rm di}$.
\item
Return to step 3 and recompute $a_{0}$.
\end{enumerate}

Note that the main computational effort is the selection
of the event (Step 5) based on $\eta_2$; to minimize the number of
operations needed to determine this we use a recursive bisection search and an efficient sorting of possible events.  Also in the recomputation of rates (Step 7) one can limit the updates to those sites that have changed and their neighbours. Finally, the computation of $a_0$ (Step 3) in subsequent steps can considerably be accelerated by using only addition and subtraction of those
rates that have changed.

\section{Simulations of phase change for GST}
\label{secResults}

For the remainder of this paper we model the phase change material GST used for read/write optical and electrical data storage devices, as in \cite{SW_2004}. Such a material has a fine balance between bulk and surface energies of crystals, meaning that one can find non-trivial nucleation and growth dynamics that varies with $T$.

Let $T_m$ be the melting temperature; if we assume that
the free energy change associated with crystallisation of a single site 
varies linearly with $T-T_m$ and the energy change associated
with change in surface $A$ is $\sigma A$ with $\sigma$ constant,
then the rate $\xi(T,A)$ can be written as
\begin{equation}\label{eq:xiT2}
\xi(T,A)~=
\exp\left[L\left( 1.0 - \frac{T}{T_m}\right)-\left( \frac{\sigma A}{k_{B}T_{m}}\right)\right].
\end{equation}
Following \cite{SW_2004}, we assume that
\begin{equation}\label{eq:L} 
L=\frac{\Delta H_{f}~v_{m}}{2k_{B}{T_{m}}}
\end{equation}
where constants are $\sigma=2.2\times 10^{-6}\, J cm^{-2}$, the interfacial 
energy density between amorphous and crystalline phases
$S_{m}=2.1187\times 10^{-14}\, cm^{2}$ the molecular surface area of the material.
We use $E_a=2.1\,eV$ and $k_0=10^{16} \mu s^{-1}$ where we use time units
of microseconds for convenience. The other constants are as follows:
\begin{itemize}
\item
$\Delta H_{f}=625~J cm^{-3}$ is the enthalpy of fusion from the data
obtained from differential scanning calorimeter experiments on GST.
\item
$v_{m}=2.9 \times 10^{-22} cm^{3}$ is the molecular volume of GST and
$S_m= 2.1 \times 10^{-14} cm^{2}$ assuming approximately
spherical shape.
\item
$T_{m}$=$889^{o}K$  is the melting temperature
\item
$k_{B}=1.381 \times 10^{-23} J/^{o} K $ is Boltzmann's constant.
\end{itemize}
Using these values we obtain $L=7.3816$. In our simulations we 
assume we have $N^2$ sites with periodic boundary conditions applied
in both directions; i.e. $r_{i+Nj}=r_{ij+N}=r_{ij}$. The parameters for the Phase Field Gillespie algorithm outlined above give realistic quantitative agreement with crystal growth in GST over a range of temperatures.

\subsection{Nucleation and crystal growth}

We simulate using an $N^2$ grid with $N=256$. Note that the crystalline fraction $X$ for such a grid can be calculated as
$$
X=\frac{1}{N^2}\sum_{(i,j)\in G} r_{ij}
$$
where clearly $0\leq X\leq 1$ and $X=1$ corresponds to a fully crystalline state.

\begin{figure}
\begin{center}
\mbox{\epsfig{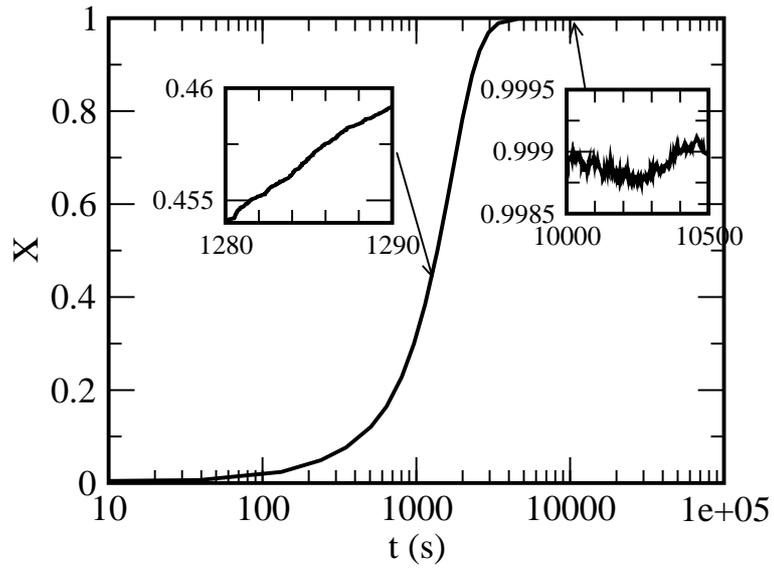}}
\end{center}
\caption{\label{fig_anneal1}
Crystalline fraction $X$ as a function of time during low temperature
anneals at $131^oC$. Detail of the progress of the anneal is shown during the growth phase and when the crystalline fraction has saturated near $X=1$.}
\end{figure}

\begin{figure}
\begin{center}
\mbox{\epsfig{file=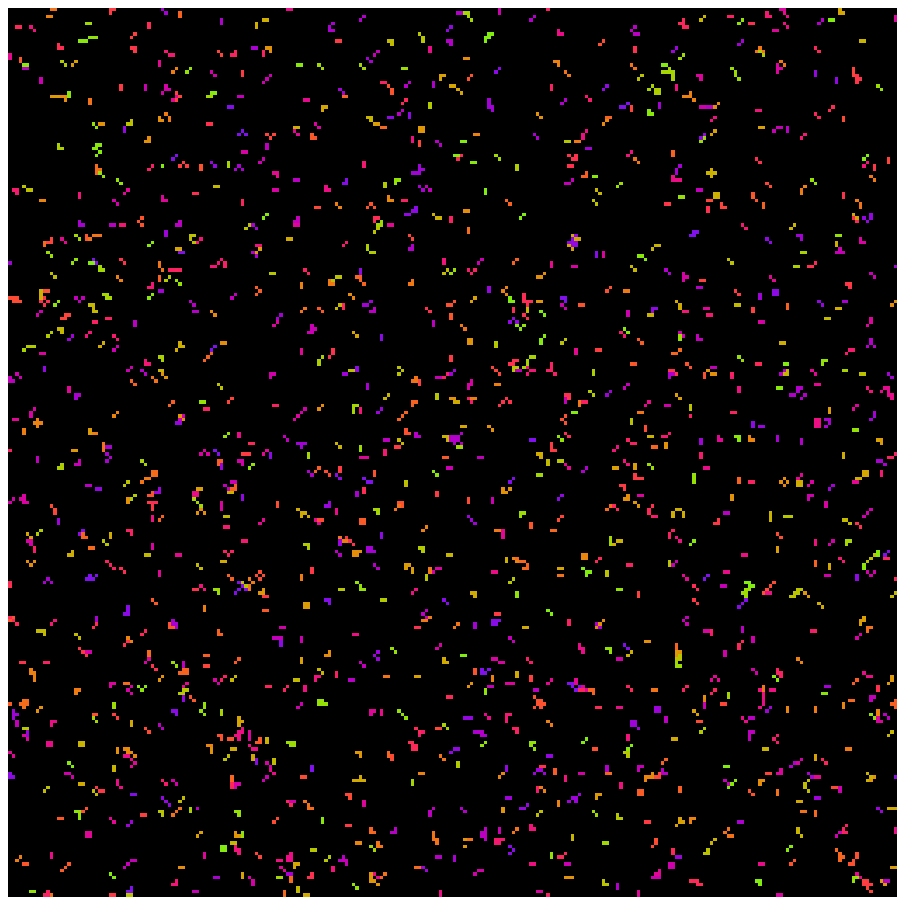,width=38mm}}
~\mbox{\epsfig{file=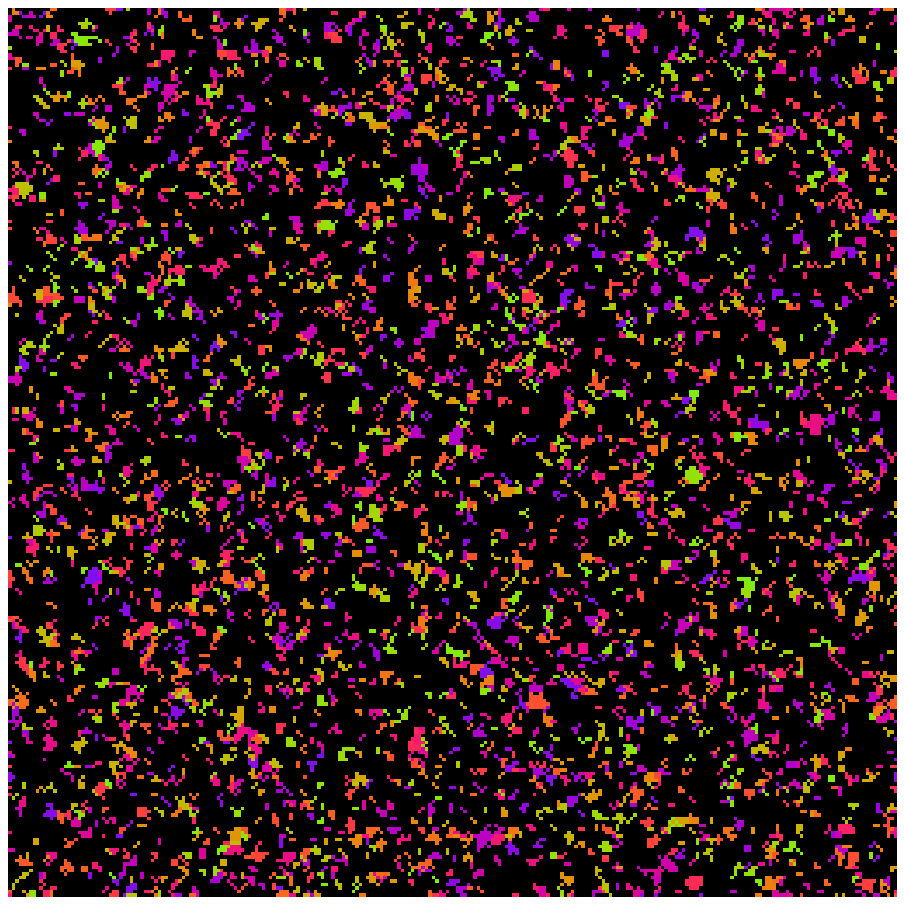,width=38mm}}
~\mbox{\epsfig{file=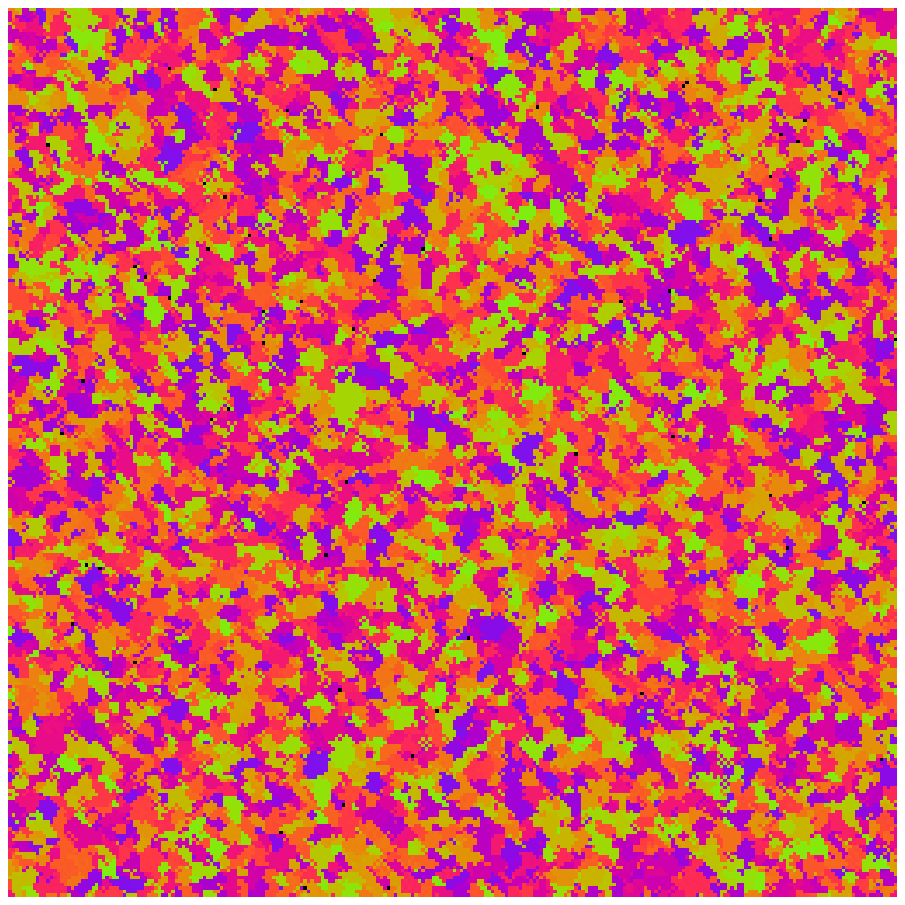,width=38mm}}

(a) \hspace{38mm} (b) \hspace{38mm} (c)

\mbox{\epsfig{file=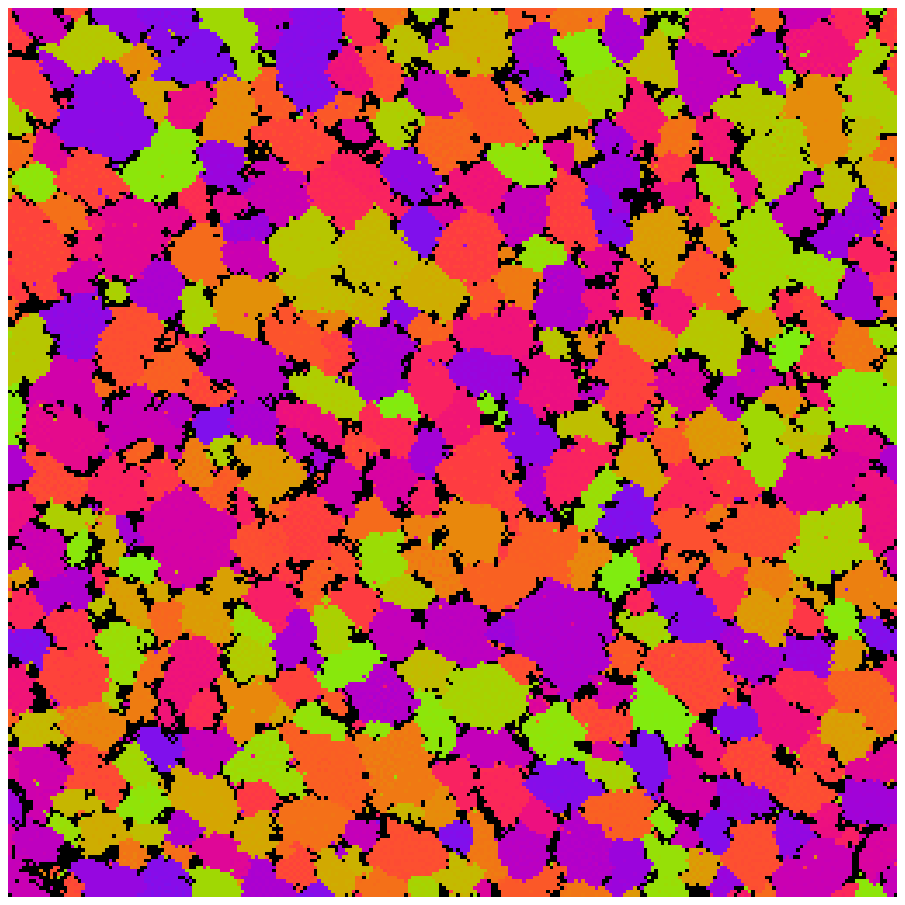,width=38mm}}

(d)

\end{center}
\caption{\label{fig_ims_131C}
Images showing progress in crystallisation (a-c) for $T=131^oC$
(as shown in Figure~{\protect \ref{fig_anneal1}}); and (d) for $T=407^oC$,
starting with pure amorphous material. The colours are assigned
arbitrarily to different oriented crystal grains. (a) shows after 2000 steps of the algorithm a number of nuclei with $X=0.0507$ after time $248s$,
(b) shows after $10000$ steps with $X=0.204$ after time $743s$,
(c) shows after $10^5$ steps with $X=0.999115$ after time $68930s$.
Similarly, (d) shows the state after $10^6$ 
steps corresponding to $4.347\mu s$ at the higher temperature.
Observe the faster progress and larger crystals that result at the higher
temperature.
}
\end{figure}

\begin{figure}
\begin{center}
\mbox{\epsfig{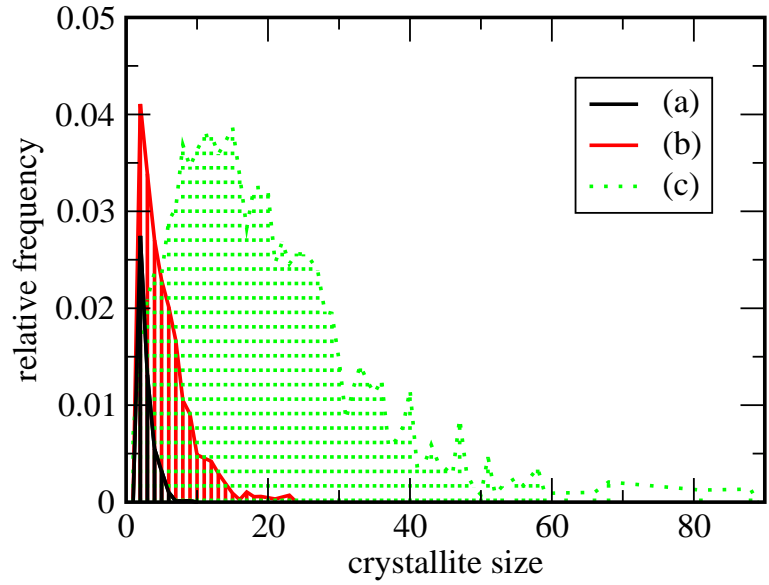}}
\end{center}
\caption{\label{fig_freqs}
The relative frequency of crystallites of different sizes corresponding to the (a), (b) and (c) of Figure~{\protect \ref{fig_ims_131C}}. Observe the peak in crystal size distribution at size 15-20 for the developed crystal structure (c).}
\end{figure}

We show in Figure~\ref{fig_anneal1} the increase in the crystalline fraction $X$ as a function of time starting at fully amorphous for $T=131^oC$; after an initial incubation the fraction quickly increases to saturate near $X=1$. The insets show that the growth occurs subject to random fluctuations  from the algorithm. Near $X=1$ there is still a nontrivial process of detachment and reattachment of sites from crystals that leads to grain coarsening over a long timescale. Figure~\ref{fig_ims_131C} shows the progress of this anneal at three stages; soon after inception, at approximately 20\% progress and in a polycrystalline state, while Figure~\ref{fig_freqs} shows the development
of the distribution of crystal sizes as the anneal progresses.

\begin{figure}
\begin{center}
\mbox{\epsfig{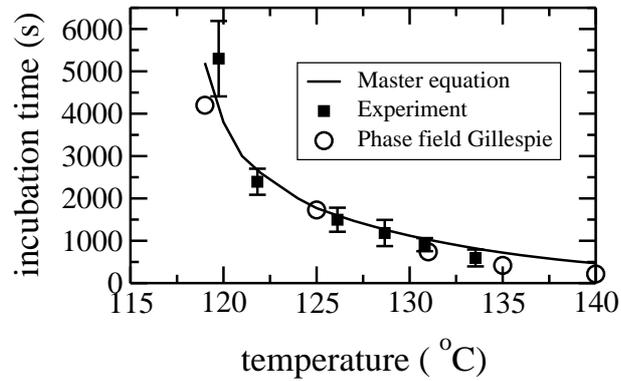}}
\end{center}
\caption{\label{fig_incubation}
Incubation times given by Phase Field Gillespie simulations, with master equation simulations of GST crystallisation from \cite{BABW_2005} and experimental data from \cite{WFZW} shown for comparison. Note that the Phase Field Gillespie simulation produces an excellent agreement with experiment.
}
\end{figure}

The incubation time (defined here as the time to 20\% crystallinity from 
fully amorphous) is shown against
temperature in Figure~\ref{fig_incubation} and for comparison the results from data from experiments
\cite{WFZW} as well as for the master equation model \cite{BABW_2005} are plotted. We note that the Phase Field Gillespie simulations show a temperature dependence that is close to experimental results of \cite{WFZW} both in form and value. As with the master
equation model, there is effectively only one fitting parameter in the model, the prefactor
$k_0$ and this is fixed independent of temperature.

%

\subsection{Spatio-temporal anneals}
\label{secComplexAnneals}

\begin{figure}
\begin{center}
\mbox{\epsfig{file=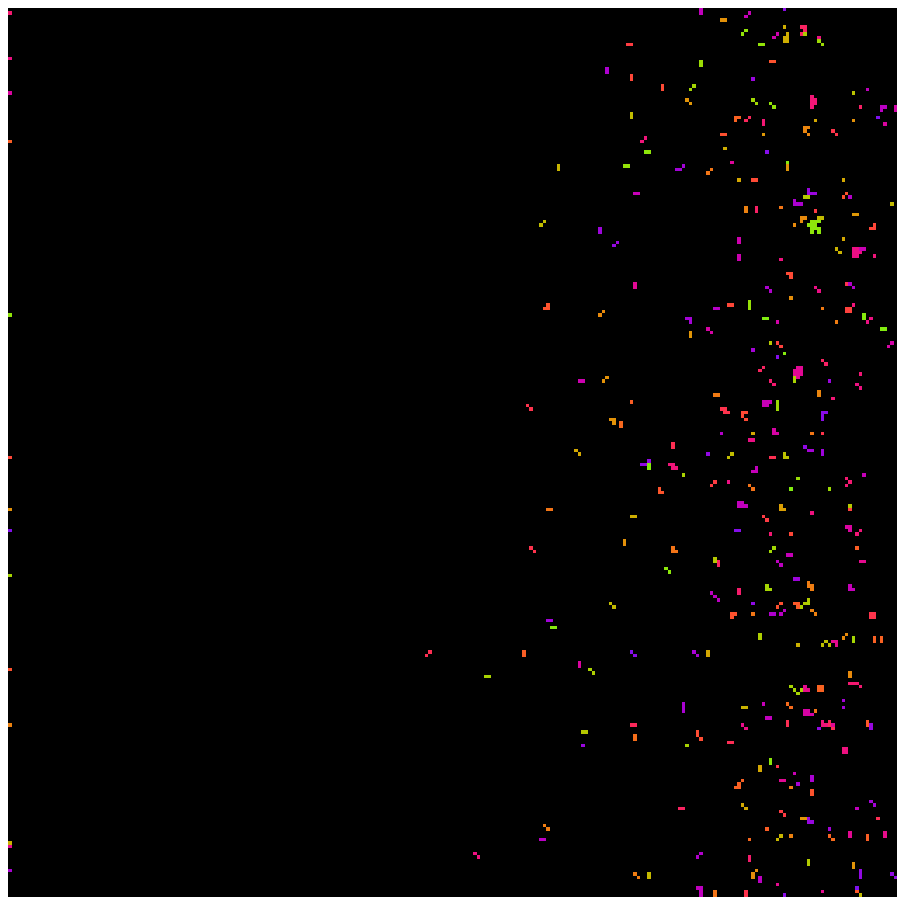,width=38mm}}
~\mbox{\epsfig{file=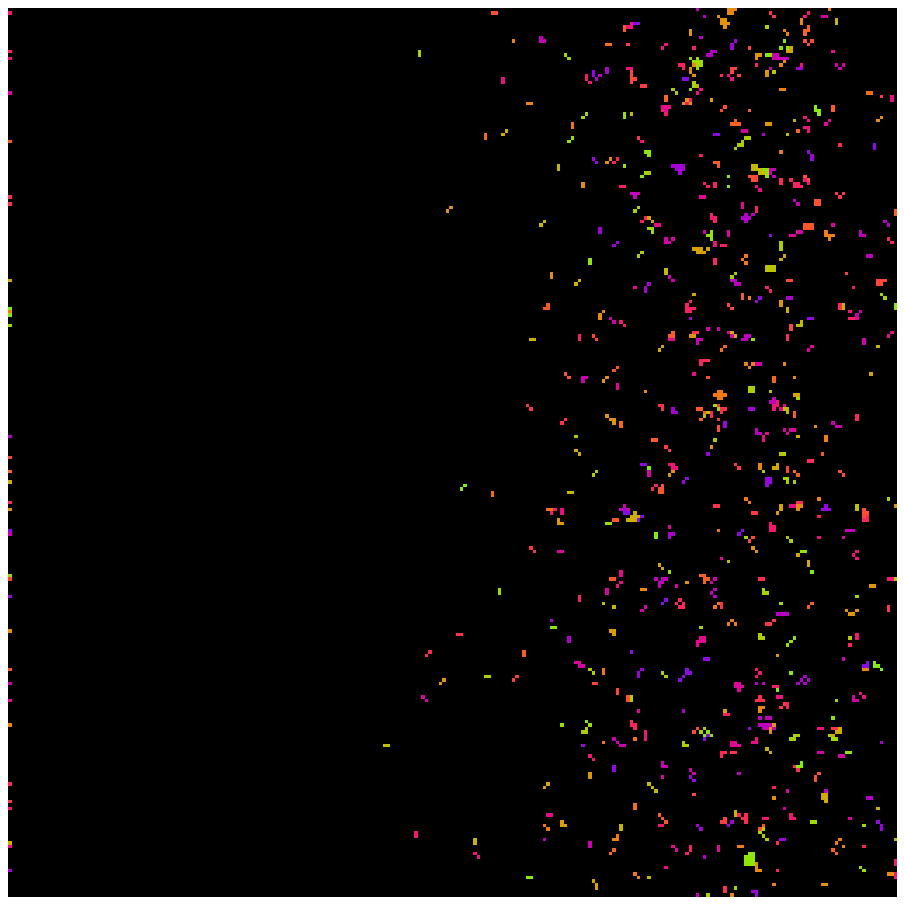,width=38mm}}

(a) \hspace{38mm} (b)

~\mbox{\epsfig{file=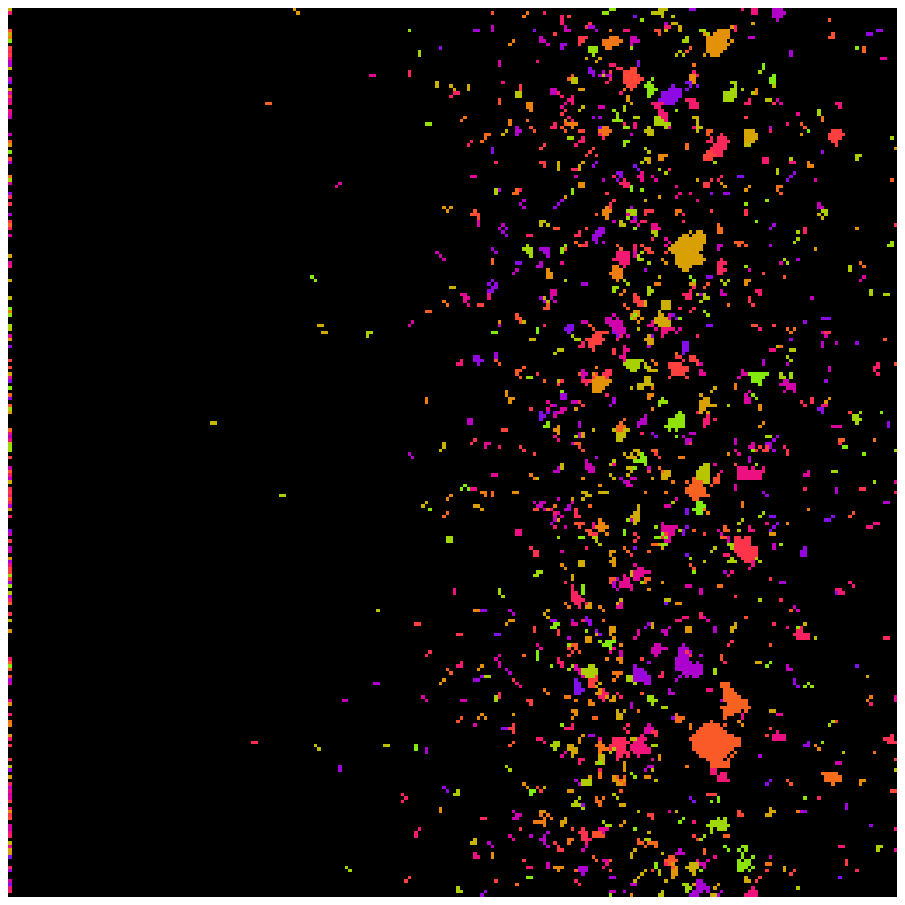,width=38mm}}
~\mbox{\epsfig{file=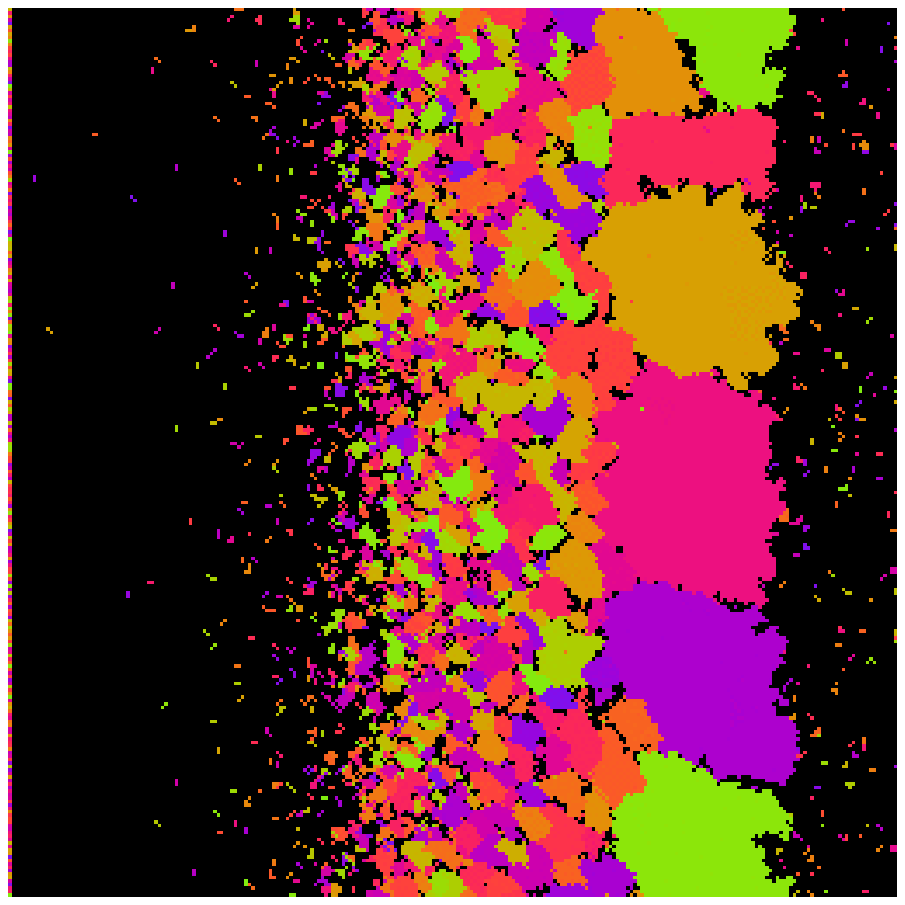,width=38mm}}

(c) \hspace{38mm} (d)

\end{center}
\caption{\label{fig_spatial}
Images showing progress in crystallisation for a sample held in
a temperature gradient where the left boundary is $227^oC$ and the right is $477^oC$.
Observe the appearence of a band of higher crystallinity as time progresses from
(a) after $17.6 ns$, (b) after $70ns$, (c) after $554ns$ and (d) after $22.9\mu s$. Observe
that the effective nucleation size is larger on the right (hotter) side of the sample.
}
\end{figure}

\begin{figure}
\begin{center}
\mbox{\epsfig{file=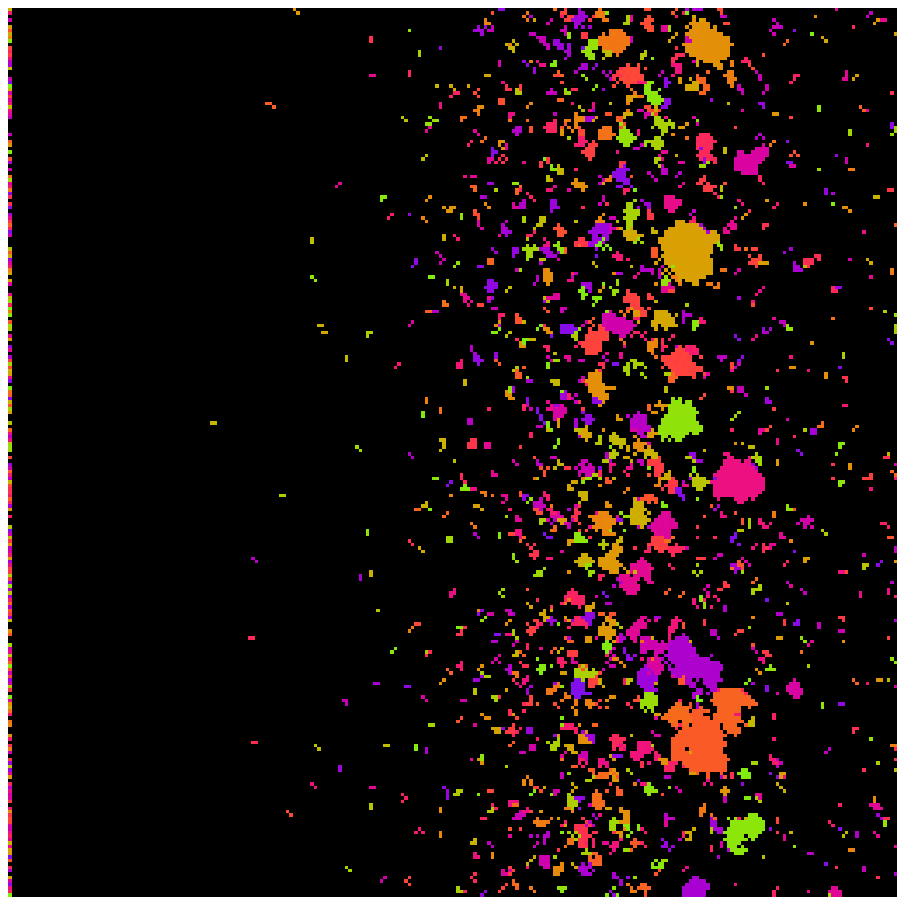,width=38mm}}
~\mbox{\epsfig{file=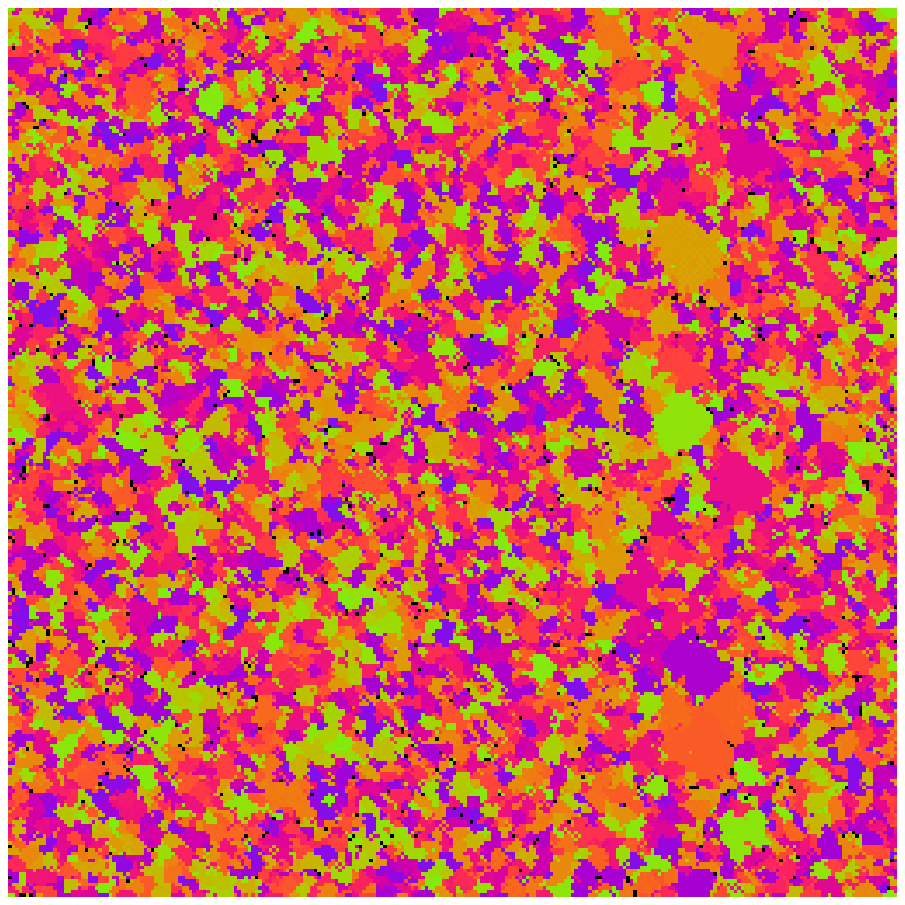,width=38mm}}

(a) \hspace{38mm} (b)

~\mbox{\epsfig{file=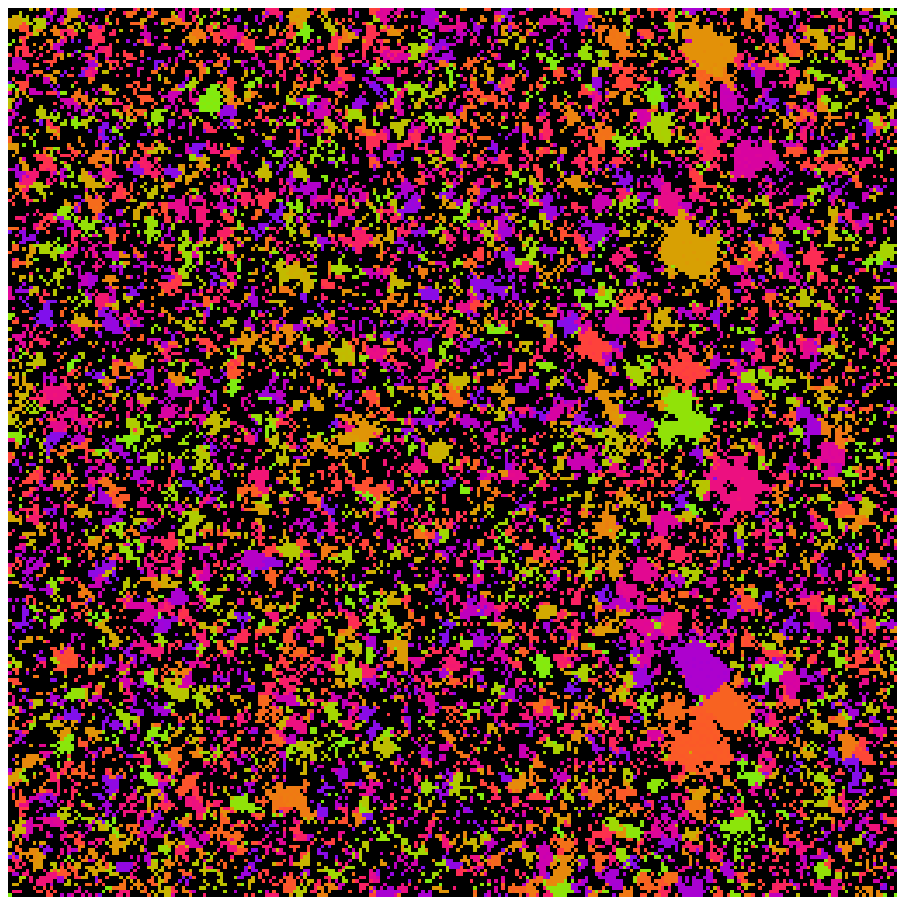,width=38mm}}
~\mbox{\epsfig{file=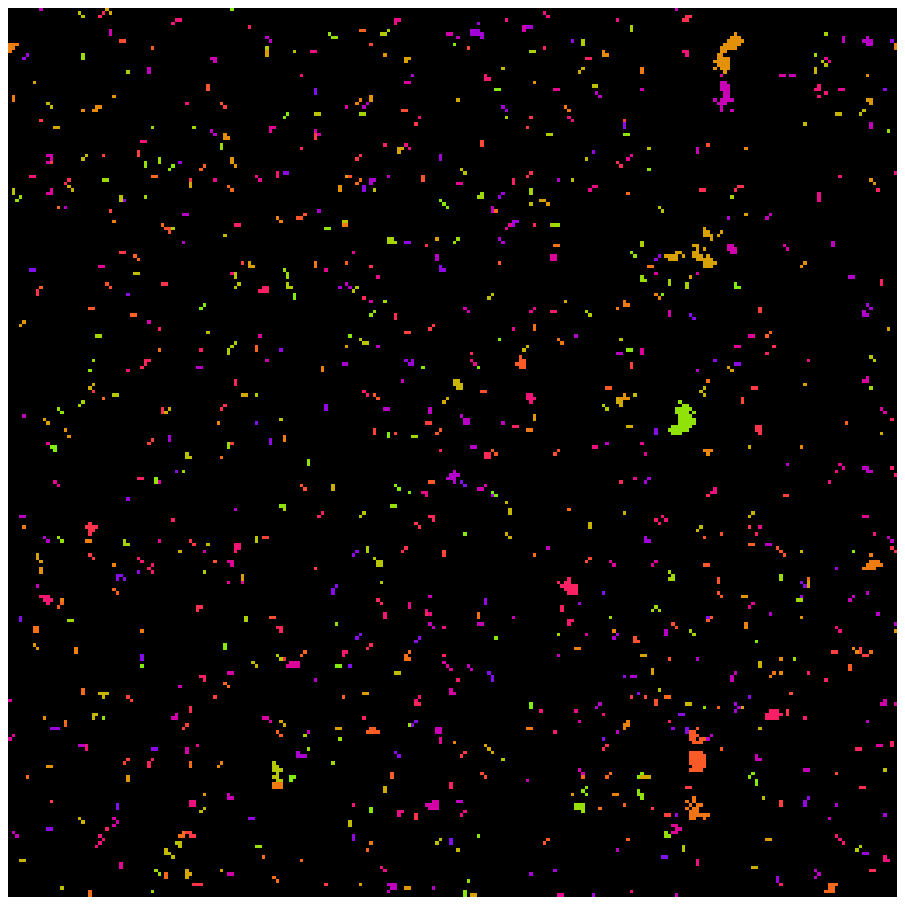,width=38mm}}

(c) \hspace{38mm} (d)

~

\mbox{\epsfig{file=fig_ms_steps.eps,width=6cm,clip=}}

(e)

\end{center}
\caption{\label{fig_spatio-temporal}
Progress of a multi-step anneal, demonstrating 
both spatial and temporal variation in temperatures.
Starting from amorphous, a sample is first subjected
$1\mu s$ of a linear spatial temperature gradient, the left at $227^o C$
and the right at $477^o C$; (a) shows resulting the crystal structure. 
For the next $0.1s$ it is maintained at $227^o C$ and in doing so 
progresses towards almost complete crystallisation but with a 
clear banded structure shown in (b). Finally the sample is 
raised to $477^o C$ for $15ns$ which is
enough for the crystals to almost entirely dissociate; the
structure at $X=0.5$ is shown in (c) and the final state
is (d). Although below melting temperature, the critical nucleus 
size is too large for crystals of this size to survive. 
(e) Shows the crystalline fraction X as a function of algorithm
step; the timesteps vary by many orders of magnitude
as the anneal progresses.}
\end{figure}

One can easily adapt the algorithm to the case where the temperature, and therefore the rates of the reactions, depends on the spatial location; the algorithm is exactly as presented before except that $T$ now depends
$T_{ij}$ on site and time. As an example, in Figure~\ref{fig_spatial} shows the development of a band of GST material that is held at $227^oC$ on the left boundary and $477^oC$ on the right boundary for a period of time. On the left hand side the growth is very slow while on the right the nucleation
energy is more difficult to overcome, meaning that initial growth is fastest in the intermediate region. A final example is given in Figure~\ref{fig_spatio-temporal} where a sample is subjected to a complex sequence of spatio-temporal anneals; see caption for details.

\section{Discussion}
\label{secDiscuss}

The Phase Field Gillespie algorithm introduced in the paper incorporates features from a few different models of crystallisation, and can be thought of as a thermodynamically
motivated caricature of a molecular simulation. We highlight a few features that could be 
investigated to make more physically realistic models:
\begin{itemize}
 \item The current model is based on a 2D grid meaning that the interpretation of the volume and
surface area of the monomers at each site should depend on interfacial energies, or alternatively 
this could be adapted to a 3D grid with suitable boundary conditions.
\item We assume the energies of the crystallites do not depend on orientation. It would be relatively easy to include anisotropy, meaning that crystallites should grow depending on their orientation.
\item We have so far only considered the behavior of the crystallisation by imposing
a temperature that may be uniform, or may have spatial non-uniformities. It would be
of interest to investigate the coupling of this to phase, for example as occurs
in electrical heating of GST. In this case onset of percolation results in lower resistance and hence increase heating.
\end{itemize}
Nonetheless the current model can evidently produce reasonably
realistic and numerically efficient simulations of crystallisation behaviour 
even for complex spatio-temporal anneals and as such we believe the model is
worthy of further investigation. We also suggest that these techniques will be useful for modelling phase change devices that use reversible transitions in GST alloys to perform computations and for multi-state storage \cite{WAB_APL,Ovshinsky_2004}.

\subsection*{Acknowledgements}
We thank Konstantin Blyuss, Andrew Bassom and Alexei Zaikin for discussions
related to this project. We also thank the EPSRC (via grant GR/S31662/01)
and the Leverhulme Trust for their support.

\newpage

%
%
%
%

\end{document}